\documentstyle[aps,floats]{revtex}
\begin{document}
\twocolumn[
\title
{Signs and Polarized/Magnetic versions of the Casimir Forces}
\author{S. Nussinov}
\address
{School of Physics and Astronomy,
Tel Aviv University, Tel Aviv, Israel\\
and Department of Physics, University of Maryland
College Park, Maryland 20742, USA}
\date{TP-95-004, quant-ph/9504019, \today}
\maketitle

\begin{abstract}\widetext
We consider  versions of the Casimir effect where
the force can be controlled by changing the angle
between two Casimir ``plates'' or the temperature of
two nearby rings.  We also present simple arguments
for the sign of Casimir forces.
\end{abstract}
\vglue0.25in
]

\narrowtext

The attractive Casimir force:\cite{Casimir}

\begin{equation}
F_{cas}= -\frac{\pi^{2}}{240}
\frac{\hbar c}{a^{4}}\frac{dynes}{cm}
\label{cforce}
\end{equation}
acts between two parallel plates at a distance
$a$ apart in vacuum.  Its independence of atomic
parameters and the QED coupling, $\alpha_E$,
reflects the perfect conductor idealization.
All details are subsumed into the boundary conditions

\begin{equation}
E_{x}=0, \, E_{y}=0, \,\,\,\, {\rm at} \,\,\,
z = 0, \,\, {\rm or} \,\, z = a,
\end{equation}
imposed on the transverse vacuum fields.
This in turn quantizes $k_{z}=\frac{n\pi}{a}$,
the z component of wave number for modes in the
region between the plates. The problem then reduces to
evaluation of the change in vacuum energy of all the
transverse modes inside this region:
\begin{eqnarray}
&\frac12 \hbar c \left\{ \int dk_x \, dk_y \,
\left( \right. \right.& \sum_n
\sqrt{k_x^2 + k_y^2 + \left(\frac{n \pi}{a}\right)^2}
\nonumber\\
&&\left. \left.- \int dk_z \sqrt{k_x^2 + k_y^2 + k_z^2}
\right) \right\}.
\label{renorm}
\end{eqnarray}
A careful regularization of this formally
divergent expression yields\cite{IZ}
\begin{equation}
E_{cas}(a) = -\frac{\pi^2 \hbar c}{720 a^3} \frac{erg}{cm^2}
\end{equation}
and $-\frac{d}{da}E_{cas}(a)=F_{cas}$ of eq. (\ref{cforce}).

The extension to spherical and other geometries involves
difficult mathematics.   Since the formal evaluation involves
delicate cancelations, often even the sign of $F_{cas}$
cannot be guessed prior to the complete calculation.
The purpose of this letter is three fold:

\begin{itemize}
\item{\bf (1)} We suggest a ``polarized'' version of the
casimir set-up in which the force is controlled by
relative rotation of the plates.

\item{\bf (2)}  We discuss the magnetic part
of the Casimir force and
speculate on its modification for superconducting rings.

\item{\bf (3)} We present general suggestive arguments
for an {it a priori} prediction of the sign
of the Casimir effect.
\end{itemize}

{\bf (1)} The two independent polarizations
$E_{x}\ne0, E_{y}=0; E_{y}\ne0, E_{x}=0$
 contribute equally to $F_{cas}$.
By an appropriate ``twist'' we can however
use these polarizations to generate a variable,
controlled,  $F_{cas}$.

The Casimir force can be also derived
by evaluating the pressure imbalance due to reflection
of vacuum modes off the outside
surfaces of the plate and of the (fewer)
internal modes off the inside surfaces\cite{Miloni}.
This derivation is completely rigorous.
It amounts to exchanging the $-\frac{d}{da}$ and
mode summation/integration.
It is however very appealing  and suggestive.  Thus,
$F_{cas}$ is almost unchanged if a
mesh of conducting wires of radius $d$ with mesh size
$b\le a$ is substituted for the plates (see fig. (1)).
The point is that the modes most relevant
for the Casimir effect those with
$\lambda\approx a$ will be equally well reflected, provided
that the ratios $a/b \equiv r_1 > 1$
 and $b/d \equiv r_2 > 1$ are not too large.\cite{comment1}

If however we have only vertical (horizontal) conducting
wires on both sides
only the $\hat{e}_{y}$ ($\hat{e}_{x}$) polarized modes will be
reflected and we readily find:

\begin{equation}
F_{cas}(parallel~~wires)=\frac{1}{2}F_{cas}
\label{paral}
\end{equation}

Next consider wires on the left and right that are crossed at
a right angle with vertical
wires on the right.  The $\hat{e}_{y}$ modes impinging
from the right on the outer part
of the right handed mesh will be exactly balanced by
 $\hat{e}_{y}$ modes impinging from the left which freely
coast through the left mesh with horizontal wires.  Thus

\begin{equation}
F_{cas}(orthogonal~~wires)=0
\label{ortho}
\end{equation}

This indeed is expected directly in the original method of derivation:
For crossed orthogonal wires, neither the $\hat{e}_{y}$ nor the
$\hat{e}_{x}$ modes are confined
to  $0 \le z \le a$.
Consequently there is no $k_{z}$ quantization and no Casimir forces.

Consider next the case when the wires
on the left make any angle
$\theta$ relative to the (say vertical) wires on the right.
The $\hat{e}_{y}$ modes impinging from the left
will now be partially reflected, with
probability $cos^{2}\theta$, from the left mesh.
This then causes
a pressure imbalance and a resulting inward
directed pressure on the right
mesh proportional to $cos^{2}\theta$.
Since Eq. (\ref{paral})
corresponds to the limiting case $\theta=0$ we have:

\begin{equation}
F_{cas}(wires~~at~~angle~~\theta) =
\frac{1}{2}cos^{2}\theta F_{cas}
\label{theta}
\end{equation}

The last equation is the most important result of this letter.
It suggests that a controlled $F_{cas}$ signal
could be expected as circular parallel rings with wires
strung along them are are rotated
with respect to each other.\cite{Sporany}

{\bf (2)} The Casimir force between two square
conducting loops of wire thickness
$\approx a$ of side $a$ a distance $a$ apart is,
on dimensional grounds:

\begin{equation}
F_{cas}(loop-loop; a, a) \approx \frac{\xi\hbar c}{a^{2}}
\label{looploop}
\end{equation}

This can be further motivated and
$\xi\approx\frac{1}{40}$ can be roughly
estimated by reconsidering the mesh of fig. (1) in the
limit $b\rightarrow a$ as follows:  since
$\lambda\approx a\ge b$ is still marginally
satisfied, there will be substantial reflection of the relevant
$\lambda\approx a$ modes.
If reflection is reduced by $\frac{1}{2}$ then the
mesh-mesh force per unit area is $\frac{1}{2}F_{cas}$.

However for $a = b$, the mesh-mesh force comes mainly from
the attraction of opposing single squares in
the two meshes, fig. (2).
Further more we can under these
circumstances approximate the mesh-mesh
force per cm$^2$, by the sum of $\frac{1}{a^{2}}$
loop-loop interactions - leading to
Eq. (\ref{looploop}).

The derivation by mode counting and
energy subtraction suggests that the
Casimir effect is equally magnetic or
electric since the electric energy and the magnetic
energy make equal contributions to
$\frac{\hbar \omega}{2}$.
None-the-less, the formal boundary
conditions are the end-products
of complicated underlying processes involving
charges and currents
induced by the $E$ and $B$ fluctuations.

Consider then two parallel conducting
rings of size $a$ and at a
distance $a$ apart.  The magnetic vacuum
fluctuations include closed
B lines which may link none of the rings,
some that link one or the other and some
which link both rings (see fig. (3)).
The last set is relevant to forces between the two rings.
It will induce, by Faraday's law, parallel
currents in the two rings.
Thus, regardless of the sign of the
B fluctuation and of the ensuing circulating current,
the resulting current
current forces will be attractive.\cite{comment2}

How will this force be modified if the rings become
superconducting?  In this case, B fields
smaller than $B_{crit}$ cannot
penetrate the superconductor if these fields
oscillate at a frequency $\omega$
lower than the critical frequency, i.e. if:

\begin{equation}
\omega \le \omega _{c} = \frac{kT_{c}}{\hbar}
\end{equation}

For high $T_{c}$ superconductors both $B_{c}$ and
$\omega_{c}$ are larger, thereby
excluding a wider range of vacuum fluctuations from
entering the superconductor.
In particular, if we put the rings at a distance $a$, such that
$a \ge \frac{c}{\omega _{c}}$,
the geometric mode cut-off in the Casimir effect,
 $\omega \le \omega_{max} \approx \frac{c}{a}$,
will automatically ensure eq. (\ref{theta}).
Also, the magnitude of the relevant
B fluctuations on this
scale $B^{2} \approx \frac{\hbar ^{2}}{c^{2}a^{4}}$
is small enough to ensure $B\le B_{crit}$.

The superconducting rings will then
impose a new important integral
constraint, namely that the total fluxes threading the various
superconducting rings must be
integer multiples of the flux quantum:
$\Phi = n\Phi _{0} = \frac{n\pi \hbar}{e}$.
Since the fluctuation of interest are of scale
$\lambda \approx a$ the above current -
current forces on the various segments of the rings add
coherently corresponding to the net current flow in the rings
$R_1, R_2$.   If there is no net global
flux change in the rings due to the vacuum
fluctuation there will be - in this approximation -
no net current and no net force.

The quantization condition implies however a strong exponential
supresion for all $n\ne 0$ sectors.  Thus if we have a
fluctuation with rougly
constant $B$ on scale $a$:

\begin{equation}
\pi B a^{2} \approx n\Phi _{0} \approx \frac{n\hbar}{e}
\end{equation}
The action of such a configuration will therefore be:

\begin{equation}
A=  \int (cB)^{2} d^{3}x dt \approx \pi^{2} c^{2}
B^{2} a^{3} \frac{a}{c} = \pi^{2}c(Ba^{2})^{2} =
\frac{cn^{2}\hbar^{2}}{e^{2}}
\end{equation}

The exponential supression will then be

\begin{equation}
e^{-\frac{A}{\hbar}} \approx e^{-\frac{n^{2}}{\alpha_{em}}}
\end{equation}
rendering such fluctuation and the attendant
magnetic Casimir forces completely negligible.

There is an amusing similarity between this supression
and that of instanton tunneling or
the probability of exciting classical configurations
with $O(\frac{1}{\alpha})$ coherent photons
around the $n=0$ sector by vacuum fluctuation,
i.e. in perturbation theory.  A particularily interesting
example is the creation of a monopole
anti--monopole pair which generates precisely
a flux $= \Phi_{0}$.\cite{Drukier}

The above considerations suggest
that if the Casimir force between
conducting rings is constantly monitored as
the temperature of the
system is lowered below the superconducting
critical temperature, then the
quenching of part of the Casimir force due to the
magnetic-current inducing fluctuation
reduces the observed force.
Even for $T_{c} \approx 100^{0} Kelvin\approx 10^{-2}ev$,
the highest superconducting temperature to date the
minimal distance
$a\ge \frac{\hbar c}{\omega_{c}}, \omega_{c}\approx 10^{-2}eV$
is $20\mu$.  At such a distance the full ordinary
Casimir force per cm is $\approx 10^{-7}$ dyne.
This is a  very small force, comparable to that
exerted on the tip of a tunneling
force microscope due to a single van-der Waals bond!

{\bf (3)} The Casimir forces can be viewed as arising from
{\bf E} vacuum fluctuations on a scale
$\lambda\approx a$ inducing opposite
sign patches of charge density
on the opposite plates (see fig. 4).
The attractive electrostatic force between such
patches yields the negative $E_{cas}$ of eq. (\ref{cforce}).

The interpretation of the Casimir energy as
the electro(magneto)static interaction of the induced
charges and currents is inspired by the original paper of Casimir
and Polder\cite{CasPol}.  The latter introduced the ``retarded''
$r^{-7}$ potentials between neutral atoms at large distances.
In the pre--Feynman--diagram nomenclature used there,
the leading $r^{-6}$ contribution of Coulomb--Coulomb
second order perturbation
cancels against part of the transverse photon--Coulomb,
third order perturbation.
The remaining part of the latter contributes the $r^{-7}$ force.
It exactly features
an induced charge distribution, say induced dipole, interacting
via Coulombic electrostatic
interactions.  This interpretation is however of heuristic
nature.  In particular, it cannot
readily account for the repulsive induced electric
magnetic--interaction\cite{Hush}.

Consider next a general ``conductor'' consisting of the union of $n$
surfaces $S_{1}, .... S_{n}$.  Let $\rho_{ind}(r,t)$ and
${\bf j}_{in}(r,t)$ with $r\in{S_{i}}$ be the induced charge
and current distributions.  The electrostatic
Casimir energy is then:
$\int d^3\,r d^3\,r'\frac{\rho_{ind}(r)\rho_{ind}(r')}{\vert r-r'\vert}$.

We can readily verify that
$E_{cas}\{S_{i}\} \ge 0$ by transforming the
electrostatic and magnetostatic energies into
$\int d^3\,r({\bf E}_{ind}^2+{\bf B}_{ind}^2)\ge0$.
Next let us consider a uniform dilation $r\rightarrow \lambda r$
sending our original set of conductors into dilated surfaces
$S_{i} \rightarrow {\lambda S_{i}}$ with dilated relative distances.
{}From dimensional arguments the casimir energy for the new
surfaces is related to that of the old set by:

\begin{equation}
E\{\lambda S_{i}\}= \frac{1}{\lambda} E\{S_{i}\}
\end{equation}

The generalized force conjugate to such a ``displacement''
$F_{\lambda} = - \frac{\partial}{\partial\lambda} E\{\lambda S_{i}\}$
is therefore always positive and the system as a whole tends to dilate.
For simple geometries such as a sphere or cylinder
this implies repulsive Casimir forces\cite{comment3}.

There is no conflict with the attraction of the two
Casimir plates - the force in question is conjugate only to the relative
separation and only the mutual interaction say
$\int d^3\,r d^3\,r' \rho_{L}(\,r)\rho_R(\,r')/\mid \,r-\,r'\mid$
with $\rho_L$ ($\rho_R$) the induced density on
left (right) plate is therefore manifest.
Had we included also the self interaction
(corresponding to the $\rho_L\rho_L+\rho_R\rho_R$ term),
positivity would be regained.
However the forces associated with the latter are
``mute'' strains inside the plate.

We can also argue on general grounds that for two objects
$A, B$, having similar shapes and
composition at a distance $R$ larger than their size $a$,
the Casimir force is attractive.
To this end we note that all Casimir--Older and Casimir
forces are, in the final analysis,
describable by two photon exchange diagrams.
The analysis of Ref. \cite{Feinberg} shows that
$V(R) \propto - \int_0^\infty \frac{\sigma(t) e^{-\sqrt{t} r}}{r} dt$
with $\sigma(t)$ the $t$--channel discontinuity of the
$A\bar{A} \rightarrow \gamma \gamma \rightarrow B \bar{B}$
amplitude.  For $A=B$, the latter is inherently positive and the
force is attractive even if $A$ and
$B$ are macroscopic objects.

I am indebted to C. K. Au, A. Casher, J. Milana and J. Sucher
for many discussions.  I have particularly benefited from and
enjoyed many most helpful discussions with Vel Hushwater.
This work was supported in part by DOE Grant DOE-FG02-93ER-40762.

\newpage
\begin{figure}
\vglue 2in
\caption{The two wire meshes replacing the two Casimir plates.
The mesh--mesh distance $a$, the mesh square size $b$, and
wire radius $d$ satisfy $a>b>d$ but with not too large $a/b, b/d$
ratios.}
\end{figure}
\begin{figure}
\vglue 2in
\caption{Two opposing squares of side $b = a$ from the two
original meshes.}
\end{figure}
\begin{figure}
\vglue 2in
\caption{Two conducting rings $R_1$ and $R_2$ (the cross--hatched
rings) and some {\bf B} flux lines.  {\bf B}$_{00}$ indicated by
broken line interlocks none of the rings.  {\bf B}$_{10}$ indicated
by one continuous line interlocks $R_2$ but not $R_1$.
{\bf B}$_{11}$ is a double closed line interlocking both $R_1$ and
$R_2$.}
\end{figure}
\begin{figure}
\vglue 2in
\caption{An {\bf E} fluctuation inducing opposite charges on the two
conducting Casimir plates.}
\end{figure}


\begin{references}
\bibitem{comment1}
Specifically we need that the average density of conduction
electrons in the mesh
$n \propto r_2^2 r_1 n_0$ with $n_0$ the density in the
plates will be large enough
so that the corresponding plasma frequency,
indicating where reflection will
 cut--off, is higher that $c/a$.
\bibitem{comment2}
There are also figure eight type configurations
in which the B field line threads the two conducting rings in opposite
directions, these would induce a repulsive interaction.  However, the
B flux cannot self intersect.  The need of spatial avoidance causes the
flux lines to be longer and the action supression to be stronger for
such a configuration.
\bibitem{comment3}
The inherent positivity of the vectorial E.M.
interactions is manifest in such
widely different areas as the stability of matter
and the positive $\Delta I = 2$
purely electromagnetic contributions to e
lementary particle masses such as
$m_{\pi^+} - m_{\pi^0}$,
$m_{\Sigma^+} - 2 m_{\Sigma^0} + m_{\Sigma^-}$,
etc.
\end{references}
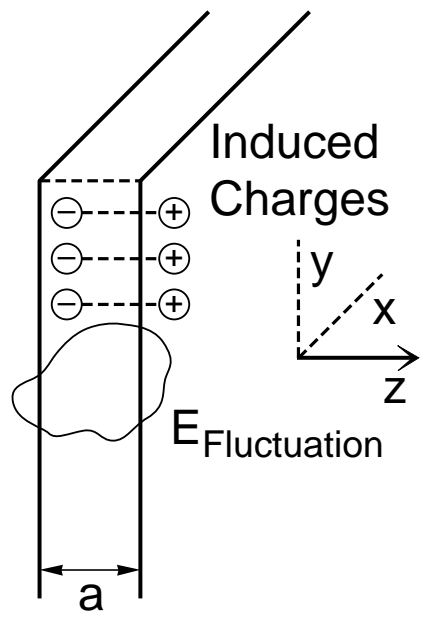
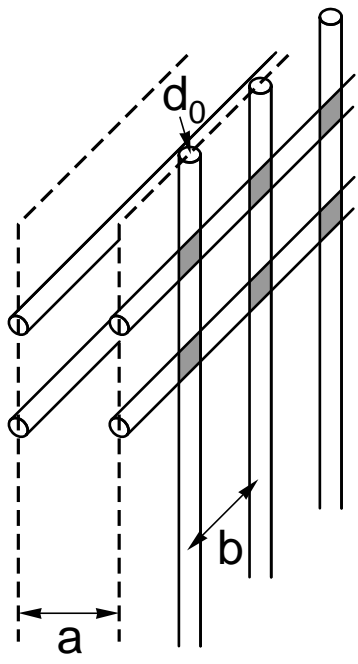
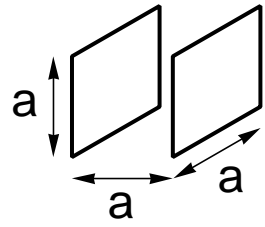
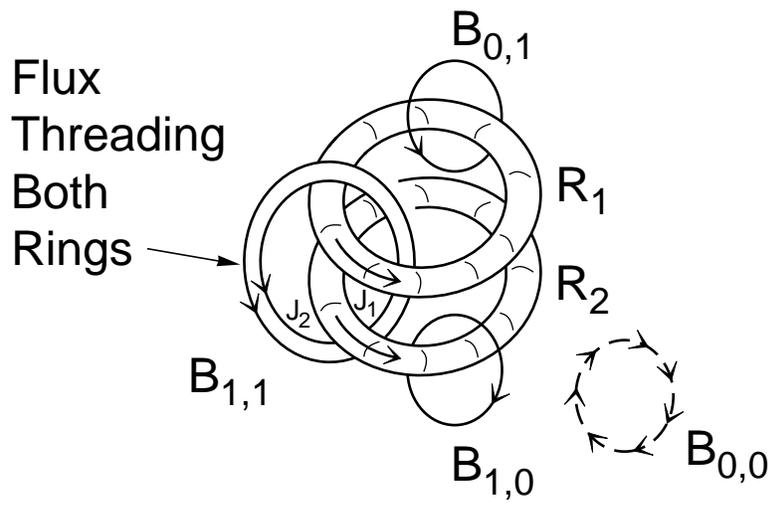
\end{document}